\documentstyle[12pt]{article}
\begin{document}
\vspace{2cm}
\centerline{\Large S.P.Novikov, A.S.Schwarz\footnote 
{S.Novikov, Math Department and IPST, University of Maryland at College Park,
MD, 20742-2431, USA and Landau Institute for Theoretical Physics, Moscow 117940, 
novikov@ipst.umd.edu, fax (USA)-301-3149363; 
A.Schwarz, Math Department,
University of California, Davis, USA. Research of the first author
is supported in part
by the NSF Grant DMS9704613}}

\vspace{0.5cm}
\centerline{\Large Discrete Lagrangian Systems on Graphs. Symplecto-Topological
 Properties\footnote
{This work already appeared in the Russia Math Surveys, 1999, v 54 n 1}}
\vspace{0.2cm}
\centerline{\Large Symplecto--Topological Properties}
\vspace{0.4cm}
This work extends the results of \cite{N1,N2,N3} where the 
Symplectic Wronskian was constructed for the linear systems on
Graphs and applied in the Scattering Theory for the Graphs with
tails.

Let us consider any one-dimensional locally finite
simplicial complex (Graph) $\Gamma$
without ends (i.e. every vertex $P$ belongs to at least two but 
to the finite number of edges
 $R_i$  only). All set of vertices is numerated by the index $j,P\rightarrow
 j(P)$. The following set of data should be given:

 1.Family $S$ of  finite subsets 
 $\alpha\in S$ in the set of all vertices. 

 2.Differentiable manifolds $N_P$ associated with every vertex $P$ 
 of our Graph $\Gamma$. Let $x_P$ be a local ccordinate in the manifold $N_P$.

 3.Smooth real-valued functions $\Lambda^{\alpha}$ 
 for every set $\alpha$; These functions are defined  on the product of all      
 manifolds $N_{j_q}$ associated with vertices belonging to the set $\alpha$
 \begin{eqnarray}
 \Lambda^{\alpha}(x_{j_1},\ldots ,x_{j_{|\alpha |}}): N_{j_1}\times \ldots
 \times N_{j_{|\alpha |}}\rightarrow R
 \end{eqnarray}

 \noindent Here $|\alpha |$ means the number of vertices belonging to $\alpha$.

 Using this data, we introduce a {\bf Nonlinear Discrete
 Lagrangian Problem:}

 {\bf The Total Lagrangian} of the system is by definition a functional $L$
 below
 defined on the functions ({\bf fields}) $\Psi (P)=x_P\in N_P$; {\bf The Euler--Lagrange 
 equation} is defined by this Lagrangian. So we have

 \begin{eqnarray}
 L\{\Psi \}= \sum_{\alpha}\Lambda^{\alpha}(x_{j_1},\ldots ,x_{j_{|\alpha |}})\\
 \frac{\delta \L}{\delta \Psi (P)}=\frac{\partial L}{\partial x_P}
 \end{eqnarray}

\newtheorem{rem}{Remark}

\newtheorem{deff}{Definition}

\newtheorem{theo}{Theorem}

We consider the Graph $\Gamma$ as a 
geodesic metric space with length of all edges
equal to 1. The {\bf Diameter} $d(\alpha)$ of any finite set $\alpha$
is equal to the maximal distance between the points of this set.

\begin{deff}
The Lagrangian Problem is called {\bf Local} if there exists a number $M$
such that $d(\alpha )<M$ for all subsets $\alpha\in S$. 
We say that this problem is presented 
in the {\bf tree-like form} if every set of vertices 
$\alpha$ is realized as a full set
of vertices of the simply-connected subgraph (tree)
$\Gamma_{\alpha}\in \Gamma$.
\end{deff}

\newtheorem{lem}{Lemma}
\begin{lem}
Every Local Lagrangian Problem can be presented in the local tree-like form.
\end{lem}
Proof of this lemma is easy. First of all, we add to the set $\alpha$
the shortest paths in 
$\Gamma$ transforming every set $\alpha$ into
the set of all vertices of the connected subgraph.
This step does not increase the diameter of our set. We extend our function
$\Lambda^{\alpha}$
to the new variables (vertices) trivially. If the new
subgraph is  nonsimply-connected, 
we start to remove edges (not vertices) from its cycles one by one, 
destroying all cycles.
Finally we are coming to the connected trees. Lemma is proved.

\begin{rem}As far as we know only nonlinear discrete time-independent
Lagrangian Systems on the discretized line
have been 
considered before 
where $\Gamma=Z, N_P=N, \Lambda^{\alpha}=\Lambda$  (see \cite{V}). 
\end{rem}

We consider now only local tree-like presented Lagrangian Systems.
For every pair of vertices $P,Q\in \alpha$ with indices $j,k$
we choose a unique 
oriented path $l^{\alpha}_{jk}\subset \Gamma_{\alpha}$ joining these vertices.
We have $\partial [l^{\alpha}_{jk}]=Q-P$ for the corresponding chains.
Let us introduce now a $C_1(\Gamma;P)$--valued 2-form on the infinite product
of all manifolds $N_P$ associated with vertices
$$N^{\infty}=\prod_jN_j$$
\noindent with local coordinats $(\bigcup_jx_j)$. Here $C_1$ means 
a linear space of all $R$-valued 1-chains on the Graph.
\begin{eqnarray}
\Omega=\sum_{\alpha}\Omega^{\alpha}=\sum_{\alpha}\frac{\partial^2
\Lambda^{\alpha}}{\partial x_j\partial x_k}[l^{\alpha}_{jk}]dx_jdx_k 
\end{eqnarray}

\begin{theo}
1.The form $\Omega$ is well-defined, i.e. for every
edge $R\subset\Gamma$ its coefficient is a well-defined $R$-valued 2-form
on the finite-dimensional manifold; 2. This form is closed $d\Omega=0$;
3.After restriction on the submanifold of 
the solutions of  Euler-Lagrange equation
this form is a 1-cycle on the Graph $\Gamma$ 
$$\delta L=0\rightarrow \partial \Omega=0$$
Therefore this form takes values in the group $H_1^{open}(\Gamma;R)$.
\end{theo}
Proof. The first part of this 
theorem follows immediately from the construction
of the form and locality. For the proof of the part 2 we need to use
the properties of the tree-like representation calculating the
coefficient of the edge $R\subset \Gamma_{\alpha}$ in 
the form $d\Omega^{\alpha}$. Let $\partial R=Q-P$ with indices $j,k$.
The subgraph 
 $\Gamma_{\alpha}$ minus $R$ is a union of 2 
 disconnected parts $\Gamma_P$ and $\Gamma_Q$
 containing separately the vertices
 $P$ and $Q$. 
 Let $T\in \Gamma_{\alpha}$ be any third vertex with index $i$
 and $T\in\Gamma_P$. The unique path $l^{\alpha}_{ij}$
 joining $T$ and $P$, does not contain
 the edge $R$. The unique path $l^{\alpha}_{ik}$
 joining $T$ and $Q$ contains the edge $R$.There are exactly two terms 
 containing
 [R]  in the 3-vertex configuration $P,Q,T$.
 Consider the coefficient of the edge [R] coming from
 the third derivatives in the calculation of $d\Omega^{\alpha}$: 
 $$(...)[R]\frac{\partial^3 \Lambda^{\alpha}} 
 {\partial x_i\partial x_j\partial x_k}=\partial_i(...)=\partial_k(...)$$
 \noindent We can see that this term in the coefficient appears exactly twice
 with opposite signs. Therefore $d\Omega^{\alpha}=0$.
 Let us point out that we did not used the Euler-Lagrange equation until now.
 For the proof of the statement 3, we observe that
 $$\partial\Omega|_P=\sum_{\alpha}\sum_k \frac{\partial^2 \Lambda^{\alpha} }
 {\partial x_j\partial x_k}dx_jdx_k$$
 \noindent where $P,Q\in\alpha$ and $j,k$ are the corresponding indices.
 Therefore we have
 $$\partial\Omega|_P=\sum_k\partial x_k\{\sum_{\alpha}d_{x_j}\Lambda^{\alpha}
 \}dx_k$$
 \noindent However, the Euler-Lagrange equation implies: 
 $$\sum_{\alpha}d_{x_j}\Lambda^{\alpha}=0$$
 Our theorem is proved.
\begin{rem}
For the paiwise interactions where all sets $\alpha\in S$
contain 2 points only, we construct a tree-like representation trivially:
Add to every set $\alpha=(P,Q)$  any
simple path $l_{PQ}$ joining these points. Extend the potentials
$\Lambda^{\alpha}(P,Q)$ trivially. 
In this case the second statement of our theorem is obvious
because all third derivatives of the 
Lagrangian $L$ along the variables of any triple
of distinct points are equal to zero.
\end{rem}

\begin{rem}
In the Appendix to the work \cite{N3} written by 
the present authors,
 the
$H_1^{open}(\Gamma;R)$-valued 2-form $\Omega$
on the space of the classical solutions
$\delta L=0$ has been constructed already. We considered a 
real selfadjoint operator $L_{\psi}$ which is a linearization 
of our problem near the solution $\psi$. Following \cite{N1}, we have
a Symplectic Scalar Product--''{\bf The Symplectic Wronskian}--
on the tangent space to the space of solutions in the point $\psi$:
$$W(\delta \psi_2,\delta\psi_1)=-W(\delta_1\psi,\delta_2\psi), L_{\psi}\delta_q\psi=0,q=1,2$$
This construction determines a $H_1^{open}(\Gamma;R)$-valued 
2-form on the space
of solutions for the Lagrangian Problem $\delta L=0$. However,
this construction is unique and canonical for the special cases only: 
for the case of nearest neighbors
and for the case of trees. 
Our proof that this construction leeds to the closed form, 
is based on  identification of 
this form with one obtained from the present construction. 
After this identification 
(which is 
easy) we see that  we need to invent the tree-like
representation of the local variational problem as it was done 
in the present work. Otherwise our construction of the form $\Omega$
can leed to the  nonclosed form.
\end{rem}

Following {\bf Partial Cases} have been considered already:

1.The case of discretized line $\Gamma=Z$-- see \cite{V}. The symplectic
form $\Omega$ in this case has been considered as an ordinary $R$-valued form. 
It agrees with our theorem because $H_1^{open}(R;R)=R$.
In this case our 2-form coinsides with the one constructed by Veselov.
However, even that is not obvious immediately.

2.The cases of linear operators on Simplicial Complex $K$ acting
in the spaces of chains of different dimensions--see \cite{N1,N2,N3}.
We proved in \cite{N3} that all these cases can be reduced 
to the operators acting on 0-cochains
 of the Graph $\Gamma$--the one--dimensional
skeleton in the first baricentrical subdivision $K^*$ of the complex $K$.
In these works beginning from \cite{N1} the properties of $\Omega$ 
as a symplectic $H_1^{open}(\Gamma,R)$-valued form have been applied to the
{\bf Scattering Theory for the Graphs with Tails.} In particular,
all unitarity properties of scattering follow from Elementary
Topology and  Symplectic Algebra.

{\bf Problem}: Investigate the form $\Omega$ for the pairwise
nonlocal interaction $\Lambda^{\alpha}(P,Q),\alpha=(P,Q)$.  
Let all manifolds
$N_P$ are equal to the sphere $S^m$ or to the same compact Lie Group 
group $G$, and the interaction potentials are also 
translationally invariant.
Especially interesting cases are following:  

1.The Graph $\Gamma$ is equal to $Z^n$. How to make an optimal choice of
the minimal paths $l_{P,Q}$? Which power of decay is necessary for the good
properties of the form $\Omega$ on the space of classical solutions? 
Is it always     of the order below?
$$\Lambda(P,Q)\sim n^{-n-2-\epsilon},n=d(P,Q)$$

2.The Graph $\Gamma$ is a  homogeneous tree with $f$ number of edges 
meeting each other in every vertex. It looks like the decay 
should be exponential like
$$\Lambda(P,Q)\sim\exp\{ad(P,Q)\}$$

\noindent with $a>a_0(m)$ where $a_0$ can be found easily. However, 
we do not know what is going to happen with our form 
after the restriction on the space 
of exact solutions.

\end{document}